\documentclass[twocolumn,aps,showpacs]{revtex4}
\usepackage{graphicx}

\begin{document}

%
%

\title{Influence of bulk inversion asymmetry on the magneto-optical spectrum of
a HgTe topological insulator}

\author{M. Pang and X. G. Wu}

\affiliation{SKLSM, Institute of Semiconductors, Chinese Academy
of Sciences, Beijing 100083, China}

\begin{abstract}

The influence of bulk inversion asymmetry in $[001]$ and $[013]$
grown HgTe quantum wells is investigated theoretically.
The bulk inversion asymmetry leads to an
anti-crossing gap between two zero-mode Landau levels in a HgTe topological
insulator, i.e., the quantum well with inverted band structure.
It is found that this is the main contribution to the anti-crossing
splitting observed in recent experimental magneto spectroscopic measurements.
The relevant optical transitions involve different subbands, but the
electron-electron interaction induced depolarization shift is found
to be negligibly small.  It is also found that the splitting of this
anti-crossing only depends weakly on the tilting angle when the magnetic
field is tilted away from the perpendicular direction to the quantum well.
Thus, the strength of bulk inversion asymmetry can be determined via
a direct comparison between the theoretical calculated one-electron
energy levels and experimentally observed anti-crossing energy gap.

\end{abstract}


\maketitle

%
%

\section{Introduction}

In recent years, topological insulators have attracted considerable
attentions for their exotic electronic properties.\cite{Zhang08Phys,
Hasan10RevModPhys, Moore10Nature}
For a HgTe quantum well, when the well width exceeding a critical value,
the system will become a two-dimensional topological insulator and
this theoretical prediction has been confirmed experimentally.\cite{Zhang06,
Konig07Science, Konig08}

An effective four-band model\cite{Zhang06, Konig07Science, Konig08}
is proposed to describe the two-dimensional topological insulator
made of a HgTe quantum well grown along the $[001]$ direction, and
many interesting aspects have been explored based on this model
system.\cite{Niu08, Liu08PRL, Schmidt09PRB, Li09PRL, Novik10PRB,
Lindner11naturephys, Chang11PRL}

The effective model predicts that, when a perpendicular magnetic field is
applied to the quantum well with inverted band structure, two zero-mode Landau
levels will cross each other and become degenerate at a critical magnetic
field.  The degeneracy can be lifted and an anti-crossing gap can appear when
one includes the effect of bulk inversion asymmetry existing in the HgTe
quantum well in constructing the effective model.  The edge states originated
from these two zero Landau levels have different dependence on its cyclotron
center coordinate.  The energy of electron-like state will bend upward, and
the hole-like state will bend downward.  Below and above the critical
magnetic field, these two zero Landau levels exchanges order in energy.
The above features of two zero Landau levels are the key to the explanation
of experimentally observed field driven insulator-metal-insulator
transition.\cite{Konig07Science, Konig08}

When there is an anti-crossing gap opening at the critical magnetic field, the edge
states will be a mixture of electron-like and hole-like states, it would
be interesting to know its dependence on the cyclotron center coordinate,
e.g., the possibility of a non-monotonic edge states dispersion.
This clearly demands a more detailed and careful study of the nature
of gap opening.  In a tight binding theoretical calculation,
the size of splitting at the critical magnetic field between two zero
Landau levels due to bulk inversion asymmetry has been estimated.\cite{Konig07Science}
More recently, experimental investigations are reported where magneto
spectroscopic measurements are performed on HgTe quantum well samples
which are grown not only in $[001]$ but also in $[013]$ directions.\cite{mag2011, mag2012}
An anti-crossing of the resonance modes versus the magnetic field is
observed and the bulk inversion asymmetry is proposed as one of the
possible origins.\cite{mag2011, mag2012}

In this paper, the influence of bulk inversion asymmetry in a HgTe
quantum well is studied within an eight-band ${\bf k}\cdot{\bf p}$
approach.  The strength of bulk inversion asymmetry enters as a parameter
to be determined.
It is found that the bulk inversion asymmetry leads to an anti-crossing gap
between two zero Landau levels as expected.  The relevant optical transitions
observed experimentally involve different subbands, therefore we have to
examine the electron-electron interaction induced depolarization shift
which is known to be important in low-dimensional quantum structures.\cite{Ando82RMP, kotthaus, tung}
It is found that, in both $[001]$ and $[013]$ grown HgTe quantum wells, the
depolarization shift is negligibly small.  We also study the effect due
to the tilting of the applied magnetic field.  It is found that the
anti-crossing gap only depends weakly on the tilting angle.
The depolarization shift remains negligible for small tilting angle.
Thus, the strength of bulk inversion asymmetry can be determined via a
direct comparison between the theoretical calculated one-electron
energy levels and experimentally observed anti-crossing energy gap,
without invoking a more complicated theory.
This can also provide a realistic parameter for the effective four-band
model when one uses it in the presence of an externally applied magnetic field.

This paper is organized as follows: In section II, the theoretical formulation
is briefly presented.  Section III contains our calculated results and their discussions.
Finally, in the last section, a summary is provided.

\section{Formulation and calculation}

The calculation of one-electron energy levels is based on the well
documented eight-band ${\bf k}\cdot{\bf p}$ approach.\cite{winkler}
For details about this method, e.g., the operator ordering, the
inclusion of a magnetic field, the influence of remote bands,
the influence of strain, and the modification due to heterojunction
interfaces, we refer to a partial list of publications and
references therein.\cite{bahder, burt, foreman, smith, zawadzki, rossler}
In our calculation, the influence of strain is included and is
found to be important quantitatively.  The quantum well is assumed
to be parallel to the $xy$ plane, and the external magnetic field
is along the $z$ direction when it is not tilted.  The valence
band of the HgTe quantum well is taken as the zero energy point.
The parameter for the bulk inversion asymmetry is denoted as $B_{\rm BIA}$
instead of $B$ in order to avoid possible confusion.\cite{bahder}
In our calculation, the axial approximation is not used.\cite{mag2011,mag2012}
The influence of bulk inversion asymmetry in a HgTe quantum well
is also discussed in Ref.{\cite{ssc}} for the zero magnetic field case.

Calculations are carried out for symmetric and asymmetric HgTe quantum
wells with Hg$_{x}$Cd$_{1-x}$Te as barriers.  Results shown are mainly
for symmetric quantum wells with $x=0.3$ barriers.  The asymmetrical
quantum well studied has a step well structure consisting of
a Hg$_{x}$Cd$_{1-x}$Te barrier, a Hg$_{y}$Cd$_{1-y}$Te well, a HgTe
well, and a Hg$_{x}$Cd$_{1-x}$Te barrier.
The band parameters used in our calculation are taken from a recent
magneto spectroscopy study.\cite{mag2012}
The bulk inversion asymmetry parameter $B_{\rm BIA}$, whose value is
not known, is taken as an adjustable parameter in order to see its
effect.

After obtaining electronic energy levels, different transition
energies can be easily calculated.  In order to make a clear
comparison with the experiment, one should know the nature of the
transition.  This is achieved by calculating the corresponding optical
transition matrix elements between two involved states.\cite{ziman, yu}
Assuming the two states are denoted as $|1\rangle$ and $|2\rangle$,
we will calculate $\pi_x=|\langle 1|(p_x+eA_x/c)|2\rangle|^2$,
and $\pi_z=|\langle 1|(p_z+eA_z/c)|2\rangle|^2$.  Two matrix
elements $\pi_x$ and $\pi_y=|\langle 1|(p_y+eA_y/c)|2\rangle|^2$
give the same information.  In the calculation of above matrix
elements, one should take into account the contribution from the
Bloch basis states,\cite{yang, liu} as the intersubband
optical transition is not fundamentally different from the
inter-band optical transition.  In previous studies,\cite{yang, liu}
the optical transition matrix element is calculated for the case
of zero magnetic field.  In the following, $\pi_x$
is shown in the unit of $m_e$eV with $m_e$ the free electron mass
in vacuum.

The depolarization field correction
is due to the electron-electron interaction
in the quantum well.  This effect can be taken into account via a
self-consistent linear response approach.  It can cause a shift to
the transition energy and a splitting
between two degenerated transitions.\cite{Ando82RMP, kotthaus, tung}
When the transition involves two states $\psi_{1}$ and $\psi_{2}$,
one may have a non-zero dynamical polarization charge density
$$ \delta n_{1,2}(z) = \int \psi^*_{1}({\bf r}) \psi_{2}({\bf r}) dxdy
\phantom{...}. $$
This charge density generates a dynamical electric field in the
direction perpendicular to the quantum well.  The corresponding
potential should be taken into account self-consistently.
When one has transitions from occupied states to empty states,
the relevant matrix element is given by $M_{\gamma,\gamma'}=
[(4\pi e^2/\epsilon_0) (eB/hc) d] F_{\gamma,\gamma'}$
with $\epsilon_0$ the dielectric constant of HgTe, $B$ the magnetic field
strength, and $d$ the quantum well width.
The dimensionless $F_{\gamma,\gamma'}$ is defined as
$$ F_{\gamma,\gamma'} = {1\over d}
   \int \left[\int^z \delta n_{1,2}(z')dz'\right]
        \left[\int^z \delta n_{1',2'}(z')dz'\right]^* dz
\phantom{...}, $$
where $\gamma$ denotes the transition between $\psi_{1}$ and $\psi_{2}$,
$\gamma'$ denotes the transition between $\psi_{1'}$ and $\psi_{2'}$.
The matrix element $F_{\gamma,\gamma}$ gives the shift of transition
energy, and $F_{\gamma,\gamma'}$ with $\gamma\ne\gamma'$ gives the
coupling between two transitions which may lead to a splitting between
degenerated transitions.\cite{Ando82RMP, kotthaus, tung}

\section{Results and discussions}

In Fig.1, the energy levels of a HgTe quantum well grown in the $[013]$
direction are shown as a function of the quantum well width $d$.
The in-plane wave vector is zero.
There is no externally applied magnetic field, and a symmetrical
quantum well is assumed.
\begin{figure}[ht]
\includegraphics[width=0.38\textwidth]{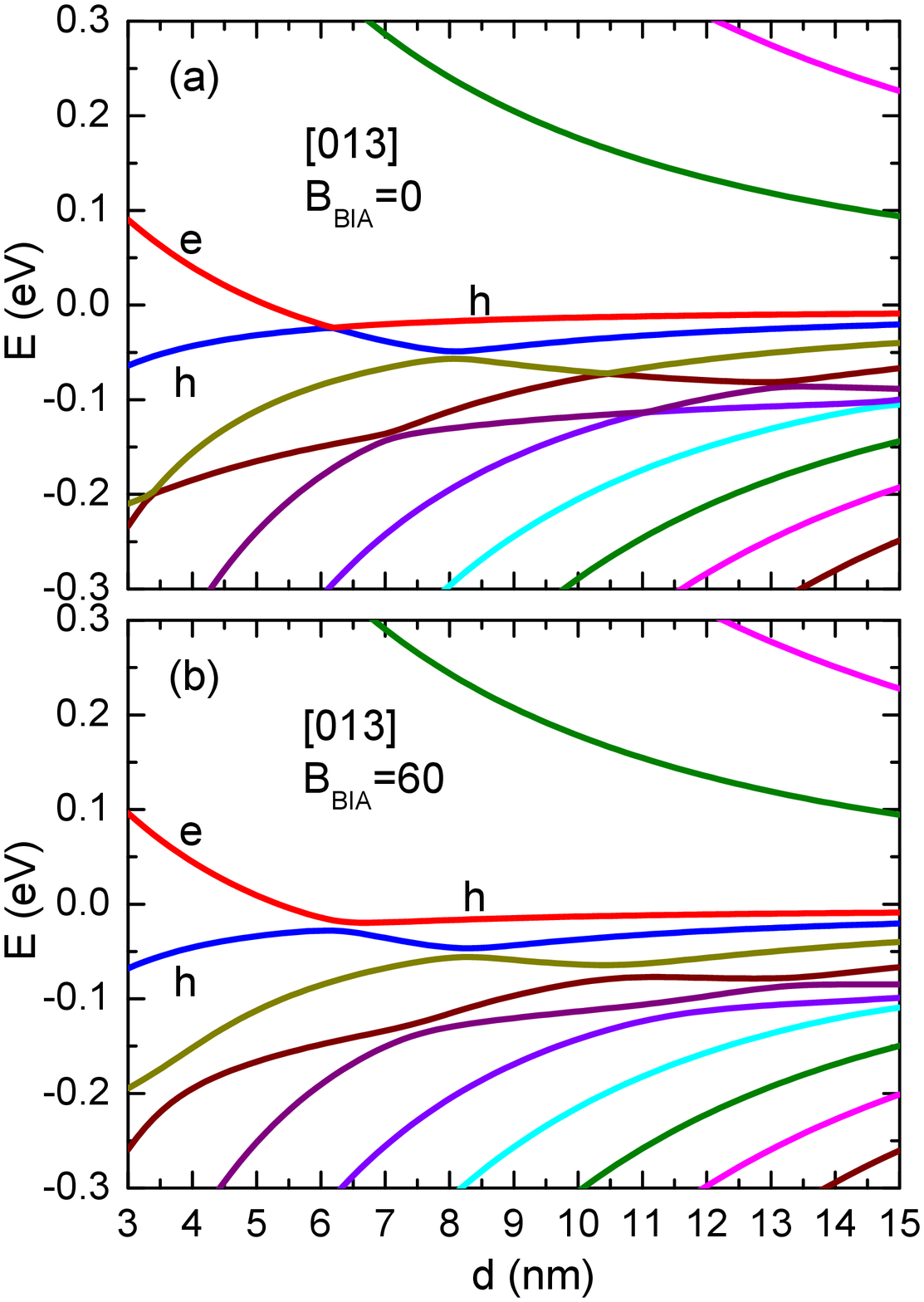}
\caption{(Color online)
Energy levels versus the quantum well width for a HgTe
quantum well grown in the $[013]$ direction, with
(a) $B_{\rm BIA}=0$ and (b) $B_{\rm BIA}=60$ eV\AA$^2$. In-plane wave vector is zero.
}
\end{figure}
In Fig.1(a), $B_{\rm BIA}=0$, and in Fig.1(b), $B_{\rm BIA}=60$ eV\AA$^2$.
Similar to the case of a $[001]$ grown HgTe quantum well, one observes that
when $d>6.2$ nm, the energy of electron-like state (marked by the symbol e)
becomes lower than the energy of hole-like state (marked by the symbol h).
One has the so-called inverted band structure.\cite{Zhang06}
Our calculation indicates that the critical well width for a $[013]$
quantum well is slightly smaller than that of a $[001]$ quantum well
of the same structure.

In the case of $[001]$ grown quantum well, our calculation indicates
that the critical quantum well width is almost not affected by the
introduction of a non-zero $B_{\rm BIA}$.  For a $[013]$ grown quantum
well, a non-zero $B_{\rm BIA}$ leads to an obvious gap opening at the
critical quantum well width.  The energies of lower subbands are also
affected.  In contrast, no obvious gap opening can be seen for a $[001]$
grown quantum well, when $B_{\rm BIA}$ becomes non-zero.  This difference
between quantum wells grown in different directions is due to the symmetry
change.  Because of the emerging of band energy order inversion, the $[013]$
grown HgTe quantum well should also be a topological insulator similar
to the $[001]$ grown HgTe quantum well.\cite{Zhang06}
\begin{figure}[ht]
\includegraphics[width=0.38\textwidth]{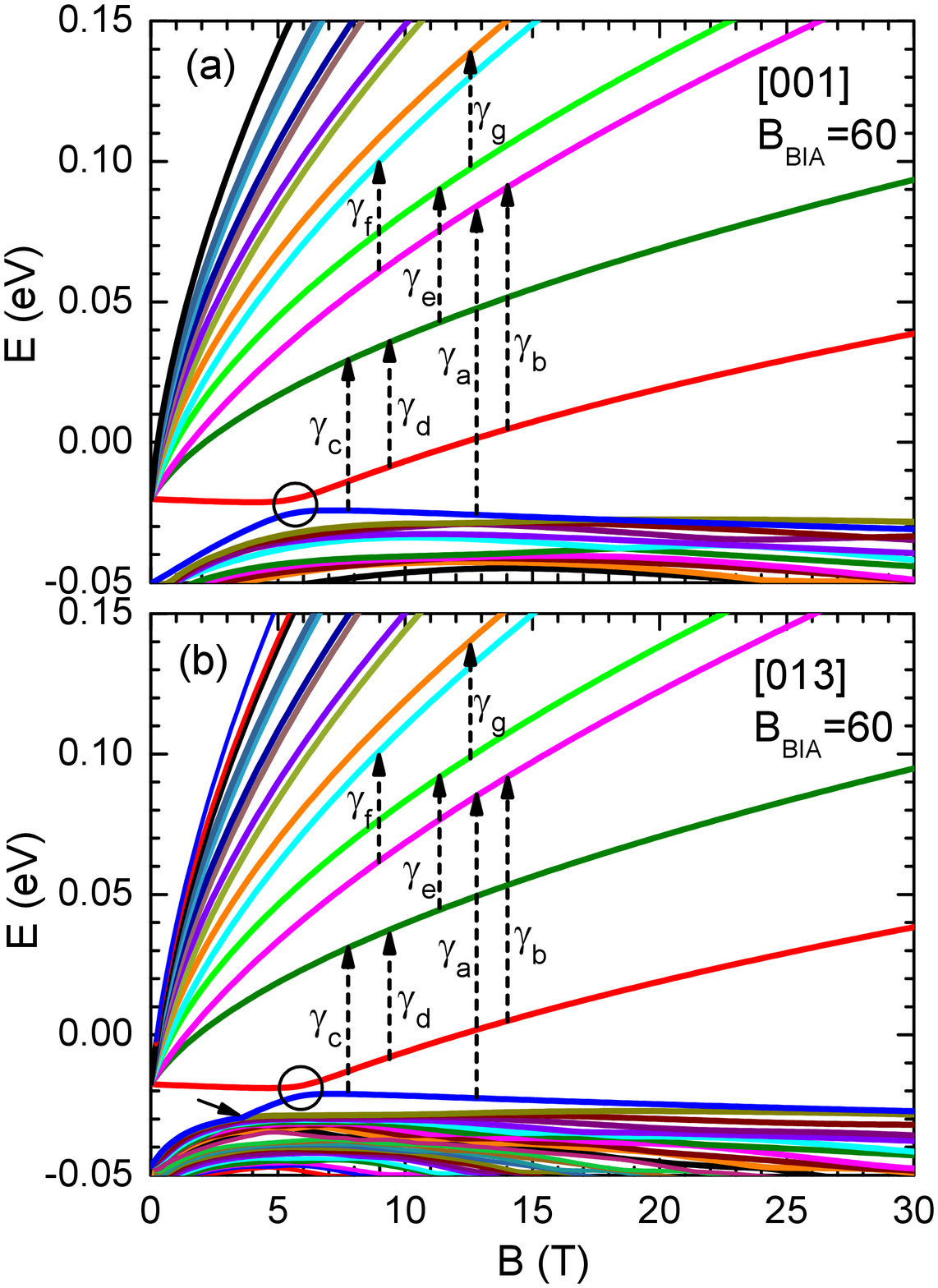}
\caption{(Color online)
Energy levels versus the strength of a perpendicular magnetic
field for $8$ nm wide symmetric HgTe quantum wells (a) grown in the $[001]$ direction,
and (b) grown in the $[013]$ direction.
}
\end{figure}

In Fig.2, energy levels are shown versus the strength of a perpendicular
magnetic field for (a) a HgTe quantum well grown in the $[001]$ direction,
and (b) grown in the $[013]$ direction.  $B_{\rm BIA}=60$ eV\AA$^2$.
The quantum well width is $8$ nm.
It is clear that an energy gap is opened between two zero Landau levels
as marked by circles.  From Fig.2(a) and Fig.2(b), one observes that the
critical magnetic fields is nearly the same for $[001]$ and $[013]$ grown
quantum wells.  In Fig.2, various transitions are labelled as $\gamma_a$,
$\gamma_b$, ..., and $\gamma_g$, with dashed-line-arrows indicating
the initial and final states involved.  In the case of $[013]$ grown
quantum well, one of the zero Landau level merges with higher Landau
levels at a non-zero magnetic field and this is marked by an arrow
in Fig.2(b).

Our calculation indicates that, when the $[001]$ or $[013]$ grown quantum
well is not symmetric about the center of the quantum well (a step well structure
with $y=0.1$), no obvious gap opening can be seen when $B_{\rm BIA}=0$.
When $B_{\rm BIA}=60$ eV\AA$^2$, it is found that, in the asymmetrical quantum well,
the gap between two zero Landau levels is not obviously changed
compared to the corresponding symmetrical quantum well.
In the reported experiments,\cite{mag2011, mag2012} the two zero Landau
levels are fully occupied, and one expects that the exchange effect due
to electron-electron interaction may shift their energy.\cite{Ando82RMP}
However, we believe that the influence of this exchange effect on the size of gap, due to
non-zero $B_{\rm BIA}$, between two zero Landau levels is small, because
the wave functions of these two zero Landau levels have similar $z$-dependence
and similar components of Landau levels.

Next, let us examine the transition energy and corresponding optical
transition matrix elements.  In Fig.3(a), the energies of various transitions
are shown versus the magnetic field.  In Fig.3(b) and Fig.3(c), the
corresponding optical transition matrix elements versus the magnetic field are displayed.  
\begin{figure}[ht]
\includegraphics[width=0.38\textwidth]{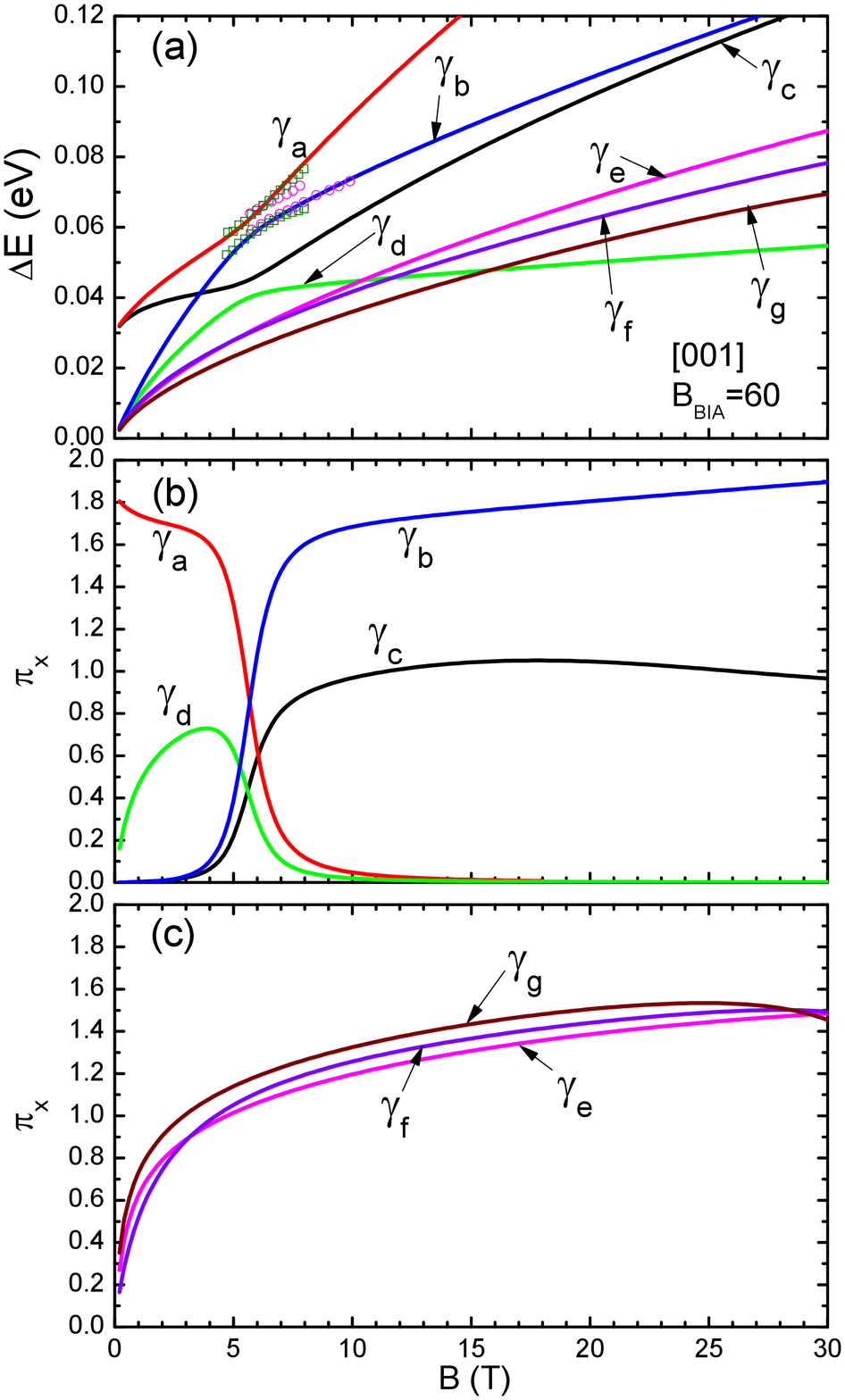}
\caption{(Color online)
(a) Energy of various transitions versus the magnetic field,
(b) and (c) The corresponding optical transition matrix elements,
for a $8$ nm wide symmetric HgTe quantum well grown in the $[001]$ direction.
Symbols are the experimental data near the anti-crossing region.\cite{mag2011,mag2012}
}
\end{figure}
The
calculation is done for a HgTe quantum well grown in the $[001]$
direction.  These transitions are selected because they are relevant
to the experiments.\cite{mag2011, mag2012}  The quantum well is a
symmetric one and the quantum well width is $8$ nm.  $B_{\rm BIA}=60$ eV\AA$^2$.
The magnetic field is perpendicular to the quantum well.
Symbols in Fig.3(a) are experimental data and this will be discussed later.

At low magnetic fields, transitions $\gamma_e$, $\gamma_f$,
and $\gamma_g$ should be observable, but they become invisible at high
magnetic fields as the initial states involved are depopulated.
The strength of these transitions is shown in Fig.3(c).  They show
similar magnetic field dependence and have similar magnitude as well.

As the magnetic field increases, transitions $\gamma_a$ and $\gamma_b$
should become observable as the final states become available for these
transitions.  Around the critical magnetic field (see also
the circle marks in Fig.2), the transition strength shows a clear anti-crossing
behavior as shown in Fig.3(b).  This is consistent with the experimental
observation.\cite{mag2011}
Transitions $\gamma_c$ and $\gamma_d$ have lower transition energies.
They also display an anti-crossing behavior but with relatively
weaker transition strength.  In an experiment, transitions $\gamma_c$
and $\gamma_d$ may not be observable if the final states involved are
fully occupied.  They should become observable at higher magnetic fields.
\begin{figure}[ht]
\includegraphics[width=0.38\textwidth]{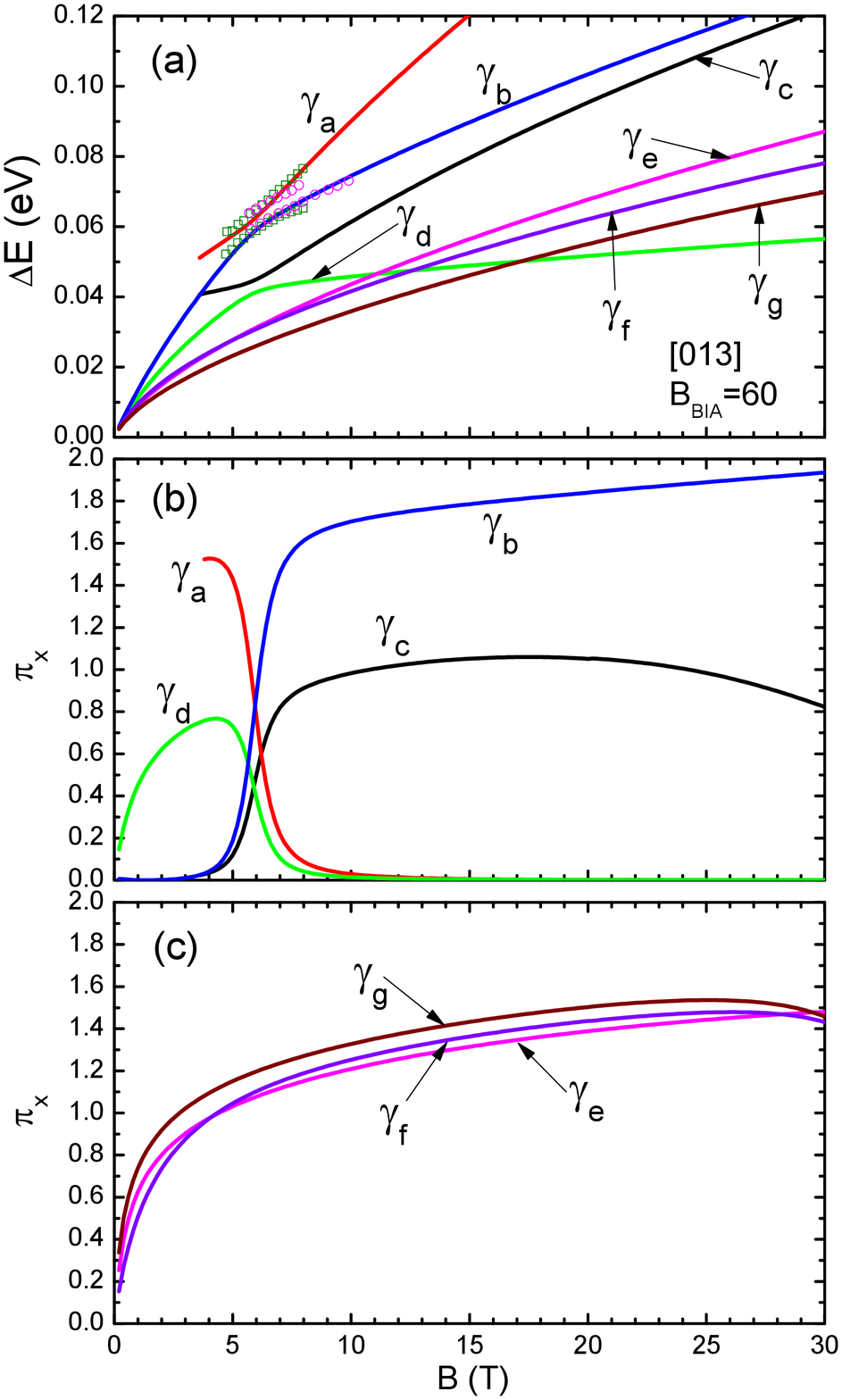}
\caption{(Color online)
(a) Energy of various transitions versus the magnetic field,
(b) and (c) The corresponding optical transition matrix elements,
for a $8$ nm wide symmetric HgTe quantum well grown in the $[013]$ direction.
Symbols are the experimental data near the anti-crossing region.\cite{mag2011,mag2012}
}
\end{figure}

We also perform calculation of the optical transition properties for a HgTe quantum well
grown in the $[013]$ direction in the presence of a perpendicular magnetic
field.  The results are shown in Fig.4.  The quantum well is a symmetric
one, and the quantum well width is $8$ nm.  $B_{\rm BIA}=60$ eV\AA$^2$.
In Fig.4(a), the energies of various transitions versus the
magnetic field are shown.  Symbols in Fig.4(a) are experimental data.
In Fig.4(b) and Fig.4(c), the corresponding optical
transition matrix elements are displayed.  One sees that the magnetic
field dependence of transition energies and of optical transition matrix
elements are qualitatively the same as that of a $[001]$ grown quantum well.
This is consistent with the $8$ nm HgTe quantum well experimental results
reported.\cite{mag2011, mag2012}

In Fig.5, the energy levels are shown as a function of $B_{\rm BIA}$, the
strength of bulk inversion asymmetry, at fixed magnetic fields and fixed
quantum well widths.  
\begin{figure}[ht]
\includegraphics[width=0.38\textwidth]{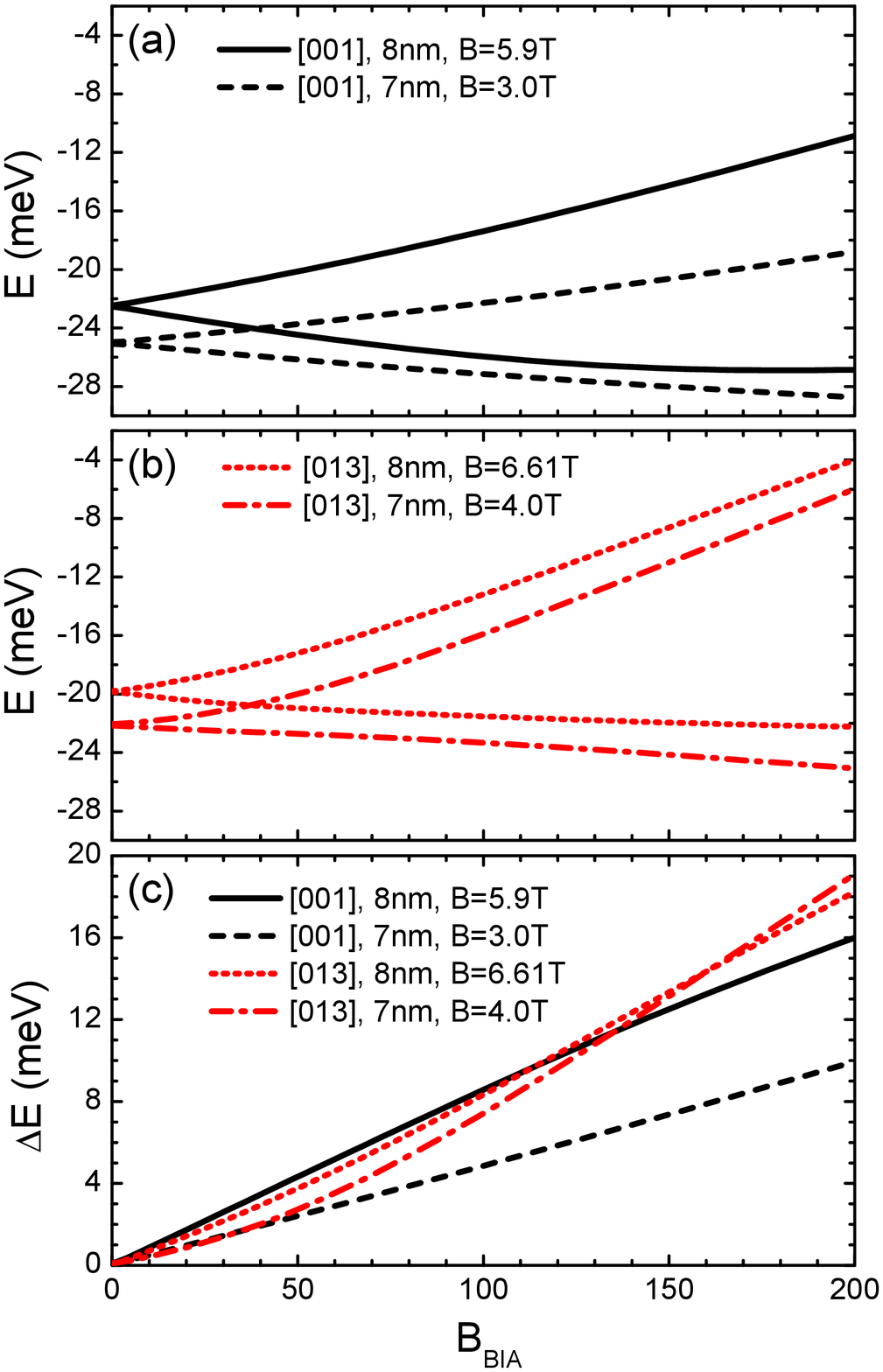}
\caption{(Color online)
Energy levels versus $B_{\rm BIA}$ at fixed magnetic fields and
quantum well widths, for (a) $[001]$ grown HgTe quantum wells,
and (b) $[013]$ grown quantum wells.  Panel (c) shows the
energy difference.
}
\end{figure}
In Fig.5(a), the result is for the $[001]$ grown HgTe
quantum wells, and Fig.5(b) is for the $[013]$ grown quantum wells.  The quantum
wells are assumed to be symmetrical ones.  In both $[001]$ and $[013]$
grown quantum wells, the critical magnetic field becomes smaller for
narrower quantum wells.  As $B_{\rm BIA}$ increases, the energy of
one state increases, while the energy of the other state decreases.
The gap between these two zero Landau levels increases as
$B_{\rm BIA}$ increases.  This is shown in Fig.5(c).  For the $[001]$ grown
quantum well, the gap size increases faster for the well of wider well
width.  For the $[013]$ grown quantum well, the gap size
evolves almost in the same way for the two wells with different well width.
The gap size for the $8$ nm $[013]$ grown quantum well can be smaller or
larger than the gap size of of the $8$ nm $[001]$ grown quantum well.
The gap size reported in the magneto spectroscopy
experiments is about $4$ to $5.5$ meV\cite{mag2011, mag2012}.
Our calculation is consistent with the experiments.

Our calculation indicates that the value of critical magnetic field
shown in Fig.2 and Fig.3 is
sensitive to the strain parameters.  In our ${\bf k}\cdot{\bf p}$
calculation, the strain is treated in the so-called coherent interface
approximation for both $[001]$ and $[013]$ grown quantum wells,
and the reconstruction of interface is not considered.\cite{decao}
Different treatment of the strain effect may produce different
quantitative results.  However, this requires a careful microscopic
structure study of the interface between quantum well and barrier
and this is beyond the scope of the present paper.

Next, we study the effect of a tilted magnetic field.
In a III-V compound semiconductor quantum well, the tilting of magnetic
field away from the perpendicular direction to the quantum well is used
to couple Landau levels and subbands or to tune the ratio between the
cyclotron resonance energy and Zeeman spin splitting.  It is a useful
experimental tool.\cite{Ando82RMP, tung}
\begin{figure}[ht]
\includegraphics[width=0.38\textwidth]{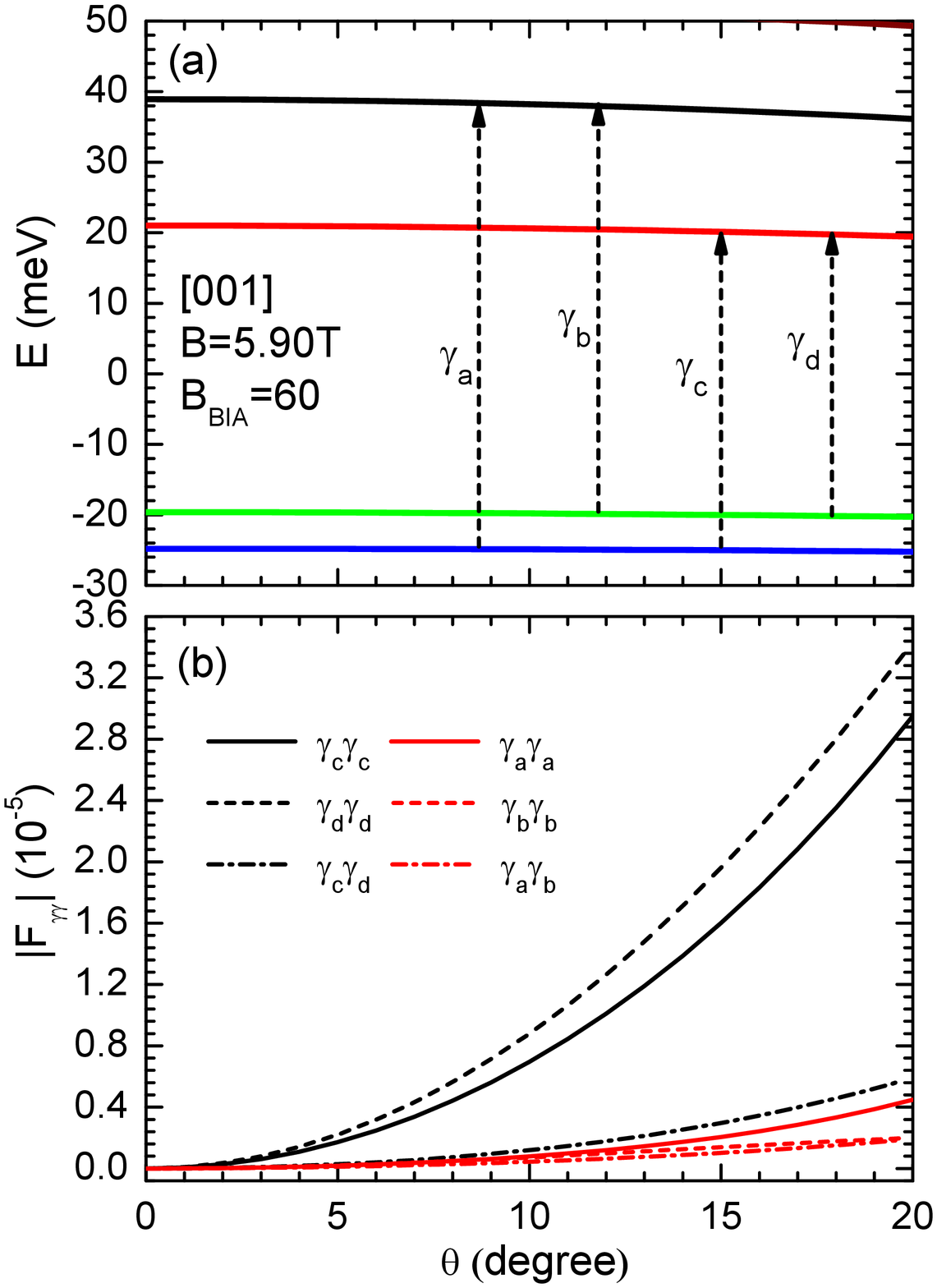}
\caption{(Color online)
(a) Energy levels versus the tilting angle and (b) depolarization
field correction versus the tilting angle for a $8$ nm wide symmetric $[001]$ grown
HgTe quantum well.
}
\end{figure}

In Fig.6(a), the energies of two zero Landau levels (two lower ones) and
energies of two higher Landau levels are shown versus the tilting
angle $\theta$.  The concerned transitions are labelled in the same way
as in Fig.2.  The calculation is done for a $[001]$ grown HgTe quantum well
of width $8$ nm.  The fixed magnetic field is $B=5.9$ Telsa,
and $B_{\rm BIA}=60$ eV\AA$^2$.  The quantum well is a symmetrical one.
One observes that two higher Landau levels show a small decrease in
energy as the tilting angle increases.  The energies of the two zero Landau
levels are almost independent of tilting angle studied.

In Fig.6(b), the depolarization field correction matrix
element $F_{\gamma,\gamma}$ is shown for various transitions versus the
tilting angle.  The off diagonal term $F_{\gamma,\gamma'}$ involving
different transitions is very small.  It is found that at $\theta=0$,
the depolarization effect can be safely ignored.
As the tilting angle increases,
the matrix elements for $\gamma_c$ and $\gamma_d$ transitions increase
more rapidly than that of other transitions.  However, for small tilting
angles shown in the figure (less that $20$ degree), the correction due to
the depolarization effect remain negligible.
The depolarization correction matrix element $M_{\gamma,\gamma'}$ is
proportional to $F_{\gamma,\gamma'}$ with a
factor $0.02(B/{\rm T})(d/{\rm\AA})$ meV for HgTe.
Thus, the correction is of the order $10^{-4}$ meV.
This value is too small to explain the experimental findings.
\begin{figure}[ht]
\includegraphics[width=0.38\textwidth]{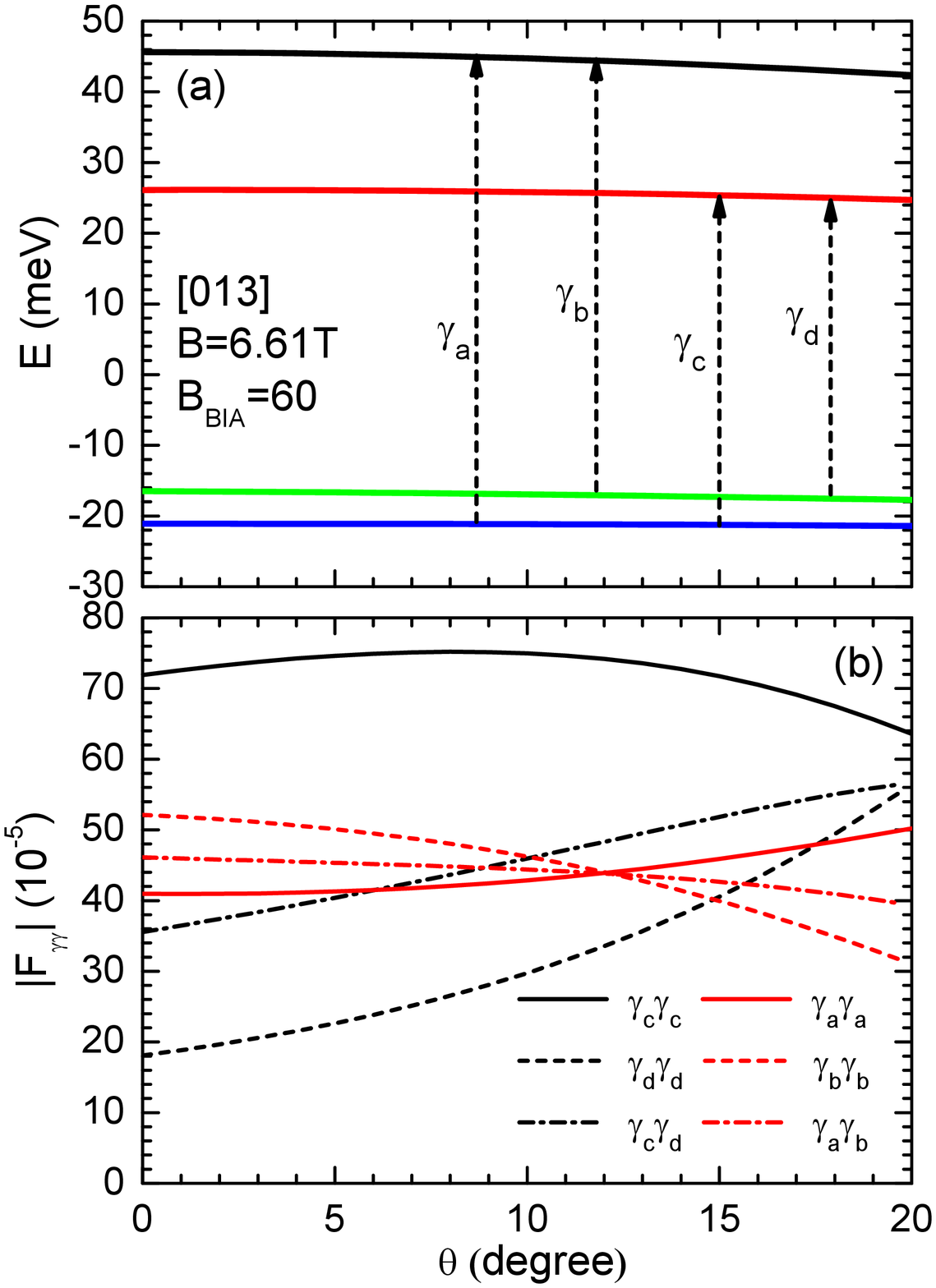}
\caption{(Color online)
(a) Energy levels versus the tilting angle and (b) depolarization
field correction versus the tilting angle for a $8$ nm wide symmetric $[013]$ grown
HgTe quantum well.
}
\end{figure}

In Fig.7(a), the energy levels of concerned transitions are shown versus
the tilting angle for a $[013]$ grown HgTe quantum well.  The quantum
well is symmetric and well width is $8$ nm.  $B_{\rm BIA}=60$ eV\AA$^2$.
One sees that the tilting angle dependence of energy levels shown in
Fig.7(a) is similar to that shown in Fig.6(a).  In Fig.7(b), the depolarization
field correction matrix element is shown versus the tilting angle.
However, different from that of a $[001]$ grown quantum well, at $\theta=0$,
the matrix elements for all transitions become obviously non-zero, and
are much larger than that shown in Fig.6(b) (see the vertical scale of Fig.7(b)).
This is due to the lowering of
symmetry in the $[013]$ grown quantum well.  However, as the magnitude
of $F_{\gamma,\gamma'}$ remains small, for small tilting angles, the
correction due to the depolarization effect is about $10^{-2}$ meV,
still too small to account for the experimental observations.

We have also investigated the tilting angle dependence for step-well asymmetric
quantum wells with $B_{\rm BIA}=0$ and $B_{\rm BIA}\ne 0$.  It is found that,
the tilting angle dependence of energy levels and depolarization corrections
is nearly the same as that shown in Fig.6 and Fig.7.  The tilting angle
dependence of concerned transition energies is small, and the depolarization
correction is negligible.  Now, we can compare our calculated results in Fig.5(c)
with the experiments,\cite{mag2011, mag2012} we estimate that $B_{\rm BIA}$
should take a value about $50$ to $60$ eV\AA$^2$.
In Fig.3(a) and Fig.4(a), experimental data near the anti-crossing
region are extracted from Refs.\cite{mag2011,mag2012} and are displayed.
The open circles are from Ref.\cite{mag2011}, and open squares are from
Ref.\cite{mag2012}.  It is clear that, the agreement is reasonably good.

%
%

\section{Summary}

In summary, the influence of bulk inversion asymmetry in $[001]$ and
$[013]$ grown HgTe/Hg$_{x}$Cd$_{1-x}$Te quantum wells is studied theoretically.
The dependence of electronic states on the quantum well width, the
magnitude of externally applied magnetic field, and the tilting angle of the
magnetic field is examined.  Our study suggests that a $[013]$ grown
HgTe/Hg$_{x}$Cd$_{1-x}$Te quantum well is also a topological insulator when the quantum
well width exceeding a critical value about $6.2$ nm.  The bulk inversion
asymmetry leads to an anti-crossing gap between two zero-mode Landau levels
and is the main contribution of this splitting.  The electron-electron
interaction induced depolarization shift is found to be negligibly small.
The splitting due to the bulk inversion asymmetry only weakly depends on the
tilting angle when the magnetic field is tilted.  Thus, the strength of
bulk inversion asymmetry can be determined via a direct comparison
between the theoretical calculated one-electron energy levels and the
experimentally observed anti-crossing splitting.

%
%

\begin{acknowledgments}

This work was partly supported NSF of China via
projects 61076092 and 61290303.

\end{acknowledgments}

%
%

%
%

\end{document}